\documentclass[cits]{PoS}

\title{Hybrid C-O-Ne White Dwarfs as Progenitors of Diverse SNe Ia}

\ShortTitle{Hybrid C-O-Ne White Dwarfs as Progenitors of Diverse SNe Ia}

\author{\speaker{Pavel Denissenkov}\\
        Department of Physics \& Astronomy, University of Victoria, P.O. Box 1700, STN CSC, Victoria, B.C., V8W 2Y2, Canada;
        The Joint Institute for Nuclear Astrophysics, Notre Dame, IN 46556, USA; NuGrid collaboration\\
        E-mail: \email{pavelden@uvic.ca}}

\author{James Truran\\
        Department of Astronomy \& Astrophysics and Enrico Fermi Institute, University of Chicago, Chicago, IL 60637 USA;
        The Joint Institute for Nuclear Astrophysics, Notre Dame, IN 46556, USA; NuGrid collaboration\\
        E-mail: \email{truran@oddjob.uchicago.edu}}

\author{Falk Herwig\\
        Department of Physics \& Astronomy, University of Victoria, P.O. Box 1700, STN CSC, Victoria, B.C., V8W 2Y2, Canada;
        The Joint Institute for Nuclear Astrophysics, Notre Dame, IN 46556, USA; NuGrid collaboration\\
        E-mail: \email{fherwig@uvic.ca}}

\author{Sam Jones\\
        Department of Physics \& Astronomy, University of Victoria, P.O. Box 1700, STN CSC, Victoria, B.C., V8W 2Y2, Canada;
        Astrophysics Group, Research Institute for the Environment, Physical Sciences and Applied Mathematics, Keele University, 
        Keele, Staffordshire ST5 5BG, UK; NuGrid collaboration\\
        E-mail: \email{swjones@uvic.ca}}

\author{Bill Paxton\\
        Kavli Institute for Theoretical Physics and Department of Physics, Kohn Hall, University of California, 
        Santa Barbara, CA 93106, USA\\
        E-mail: \email{paxton@kitp.ucsb.edu}}

\author{Ken'ichi Nomoto\\
        Kavli Institute for Physics and Mathematics of the Universe (WPI), The University of Tokyo, Kashiwa, Chiba 277-8583, 
        Japan; Hamamatsu Professor\\
        E-mail: \email{nomoto@astron.s.u-tokyo.ac.jp}}

\author{Toshio Suzuki\\
        Department of Physics, College of Humanities and Sciences, Nihon University, Sakurajosui 3-25-40, Setagaya-ku, 
        Tokyo 156-8550, Japan\\
        E-mail: \email{suzuki@phys.chs.nihon-u.ac.jp}}

\author{Hiroshi Toki\\
        Research Center for Nuclear Physics, (RCNP), Osaka University, Ibaraki, Osaka 567-0047, Japan\\
        E-mail: \email{toki@rcnp.osaka-u.ac.jp}}

\abstract{When carbon is ignited off-center in a CO core of a super-AGB star, 
its burning in a convective shell tends to propagate to the center. 
Whether the C flame will actually be able to reach the center depends on 
the efficiency of extra mixing beneath the C convective shell. 
Whereas thermohaline mixing is too inefficient to interfere with the C-flame propagation, 
convective boundary mixing can prevent the C burning from reaching the center. 
As a result, a C-O-Ne white dwarf (WD) is formed, after the star has lost its envelope. 
Such a ``hybrid'' WD has a small CO core surrounded by a thick ONe zone. 
In our 1D stellar evolution computations the hybrid WD is allowed to accrete C-rich material, 
as if it were in a close binary system and accreted H-rich material from its companion 
with a sufficiently high rate at which the accreted H would be processed into He 
under stationary conditions, assuming that He could then be transformed into C. 
When the mass of the accreting WD approaches the Chandrasekhar limit, we find a series of 
convective Urca shell flashes associated with high abundances of $^{23}$Na and $^{25}$Mg. 
They are followed by off-center C ignition leading to convection that occupies 
almost the entire star. To model the Urca processes, we use the most recent well-resolved data  
for their reaction and neutrino-energy loss rates. Because of the emphasized uncertainty of 
the convective Urca process in our hybrid WD models of SN Ia progenitors, 
we consider a number of their potentially possible alternative instances 
for different mixing assumptions, all of which reach a phase of explosive C ignition, 
either off or in the center. Our hybrid SN Ia progenitor models have much lower C to O 
abundance ratios at the moment of the explosive C ignition than their pure CO counterparts, 
which may explain the observed diversity of the SNe Ia.}

\FullConference{XIII Nuclei in the Cosmos,\\
		7-11 July, 2014\\
		Debrecen, Hungary }

\begin{document}

\section{The C flame quenching by convective boundary mixing in super AGB stars}

Denissenkov et al. \cite{denissenkov:13} have found that convective boundary mixing at the bottom of 
the C-burning convective shell prevents the C flame from reaching the center of the CO core 
in a super AGB star (Fig.~1). As a result, ``hybrid'' C-O-Ne cores are formed, 
with small CO cores surrounded by thick ONe zones, in super AGB stars with initial masses 
in the interval $M_\mathrm{u}\leq M_\mathrm{i}\leq M_\mathrm{u}+\Delta M_\mathrm{u}$ (Fig.~2, left panel). 
The pure CO and ONe cores are still formed in stars 
with $M_\mathrm{i}\leq M_\mathrm{u}$ and $M_\mathrm{i}\geq M_\mathrm{u}+\Delta M_\mathrm{u}$, respectively. 
The exact values of $M_\mathrm{u}$ and $\Delta M_\mathrm{u}$ depend on model assumptions, 
such as the initial chemical composition, carbon burning rate \cite{chen:14}, 
and efficiency of convective boundary mixing at the boundaries of H and He convective cores 
(compare the values of $M_\mathrm{i}$ in Fig.~1 and in the left panel of Fig.~2).

\begin{figure}
\includegraphics[width=.5\textwidth]{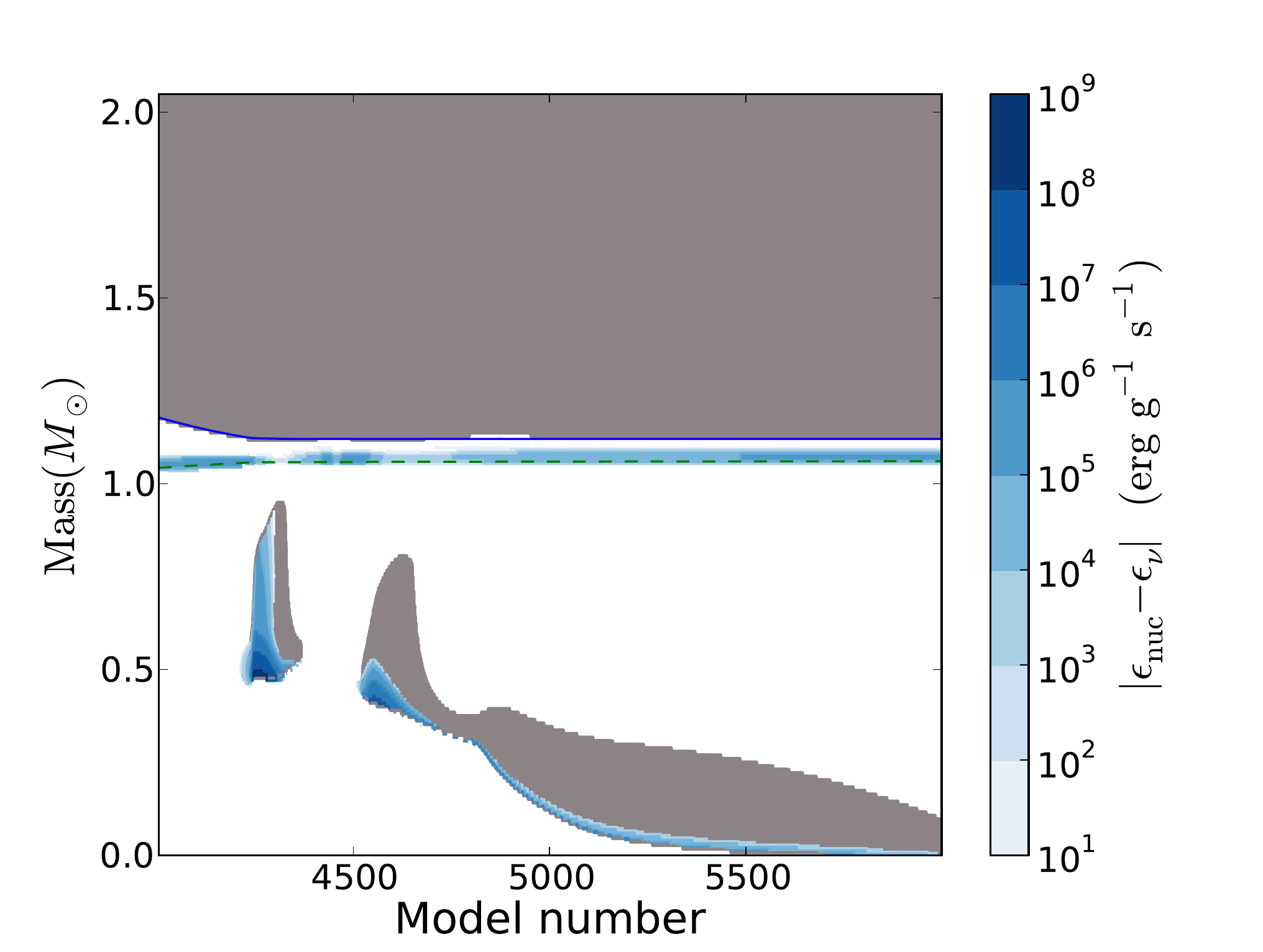}
\includegraphics[width=.5\textwidth]{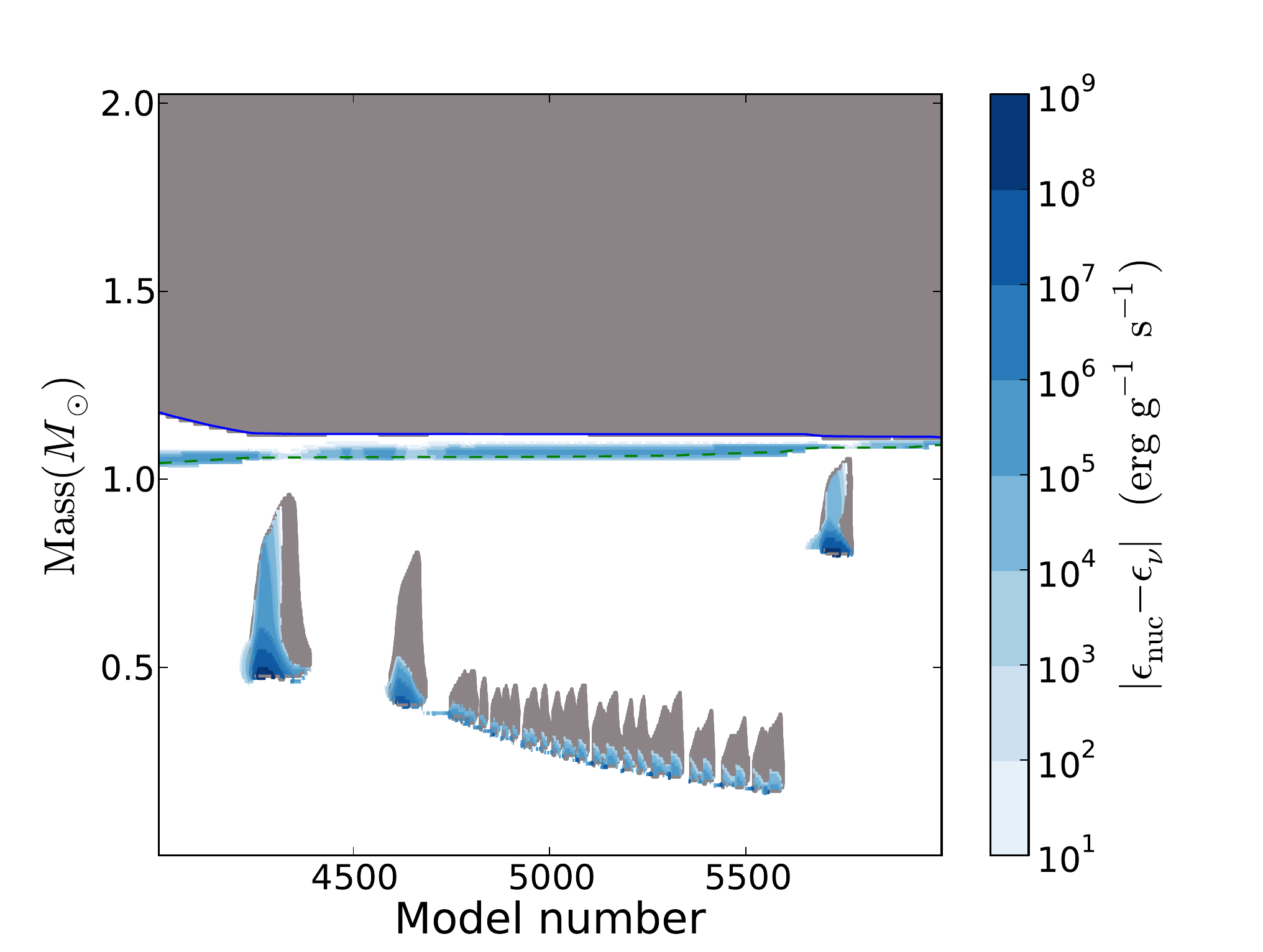}
\caption{Kippenhahn diagrams (grey regions show convective zones, blue shadows give nuclear energy 
generation rates) for a central part of a super AGB star model with the initial mass $M_\mathrm{i} = 9.5 M_\odot$
and solar composition. {\bf Left panel:} a standard case without extra mixing. {\bf Right panel:} a case 
with convective boundary mixing modeled with a diffusion coefficient that decreases exponentially 
with a distance from the bottom of the C convective shell where its value is calculated using 
a mixing length theory. Convective boundary mixing with the length scale of its exponential decay 
equal to $f=0.007$ pressure scale heights was included only during C burning, but it was neglected, 
assuming $f=0$, on the preceding phases of convective H and He core burning. The latter explains 
why $M_\mathrm{i}$ in these two cases is larger than in the left panel of Fig.~2.}
\label{fig1}
\end{figure}

\begin{figure}
\includegraphics[width=.5\textwidth]{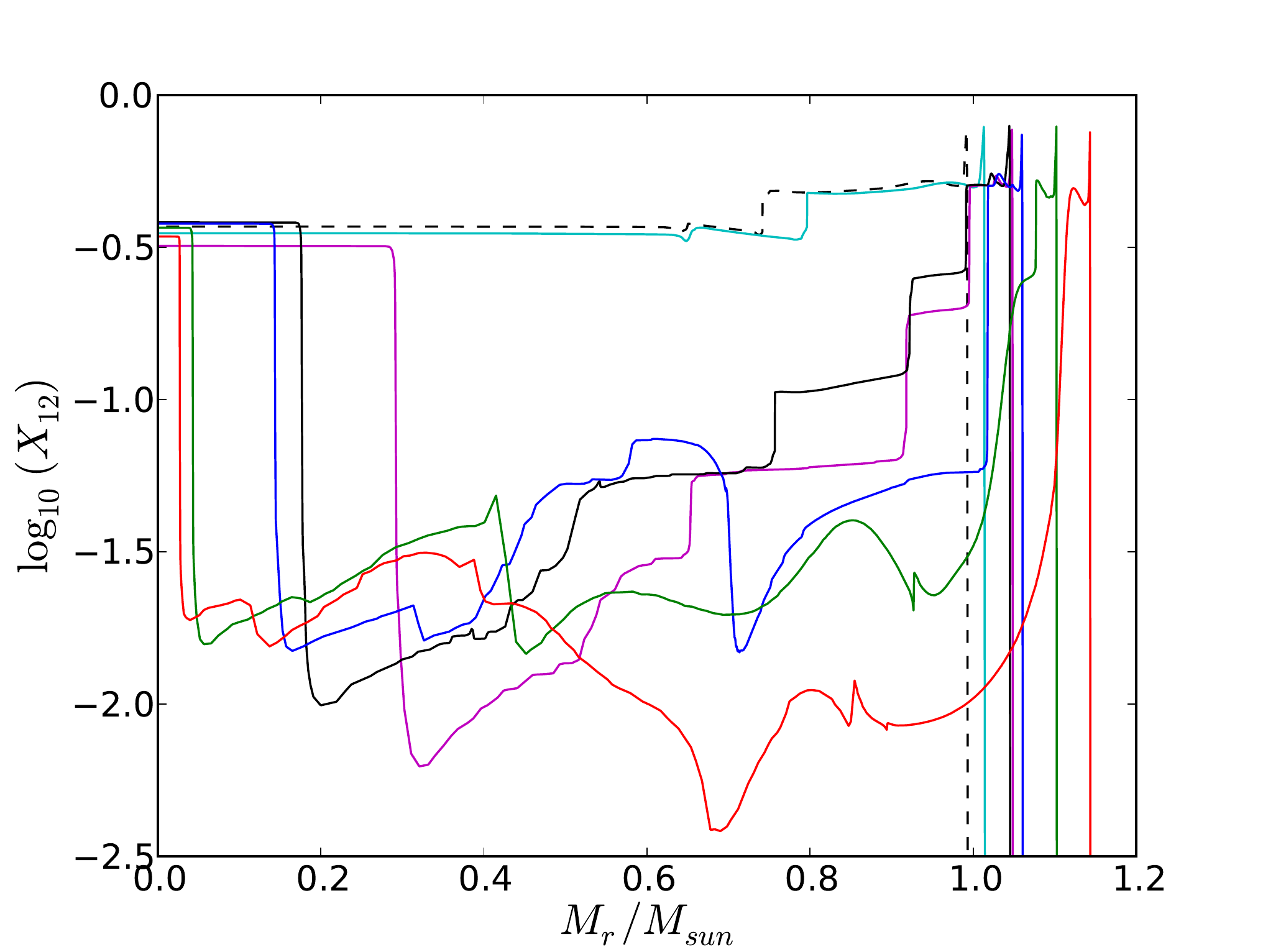}
\includegraphics[width=.5\textwidth]{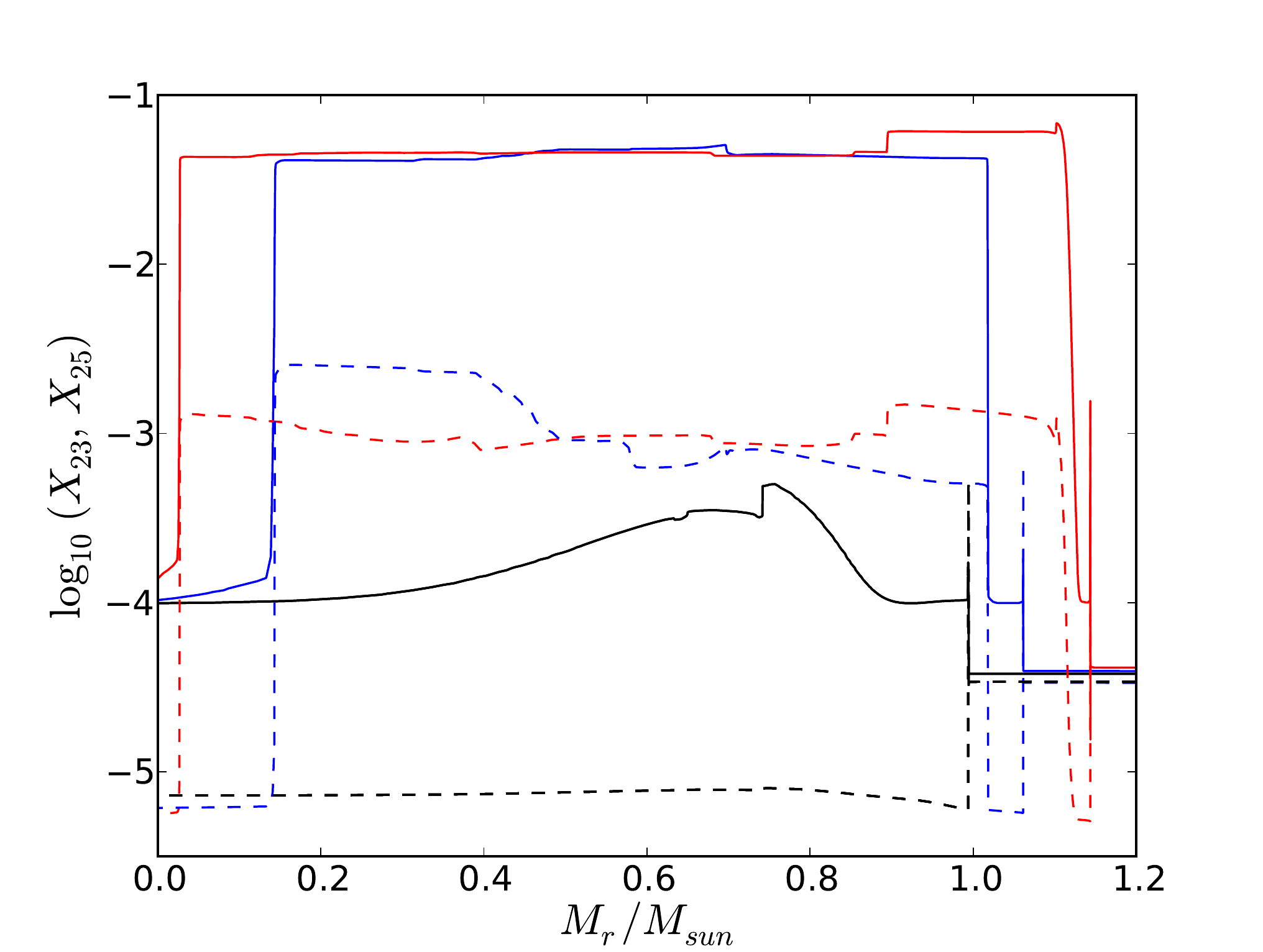}
\caption{{\bf Left panel:} Mass fraction profiles of $^{12}$C at the end of carbon burning in the hybrid C-O-Ne cores of 
super-AGB stars. A value of $f = 0.014$ was used to model convective boundary mixing at all convective 
boundaries, except the stiffer boundaries of the He and C convective shells, where a value of 
$f = 0.007$ was assumed. The initial masses of the AGB stars are $6.3 M_\odot$ (dashed black curve), 
$6.4 M_\odot$ (cyan curve), $6.5 M_\odot$ (magenta curve), $6.8 M_\odot$ (solid black curve), $6.9 M_\odot$ (blue curve), 
$7.1 M_\odot$ (green curve), and $7.3 M_\odot$ (red curve). In the first two models, C has not  
ignited in the CO cores. The initial mass fractions of hydrogen and heavy elements are $X=0.70$, $Z=0.014$.
{\bf Right panel:} Mass fraction profiles of $^{23}$Na (solid curves) and $^{25}$Mg (dashed curves)  
in the CO core of the AGB star with $M_\mathrm{i} = 6.3 M_\odot$ (black curves) after the first He-shell thermal pulse,  
and in the hybrid C-O-Ne cores of the super-AGB stars with the initial masses $6.9 M_\odot$  (blue curves) 
and $7.3 M_\odot$ (red curves) at the end of C burning.}
\label{fig2}
\end{figure}

\section{Can the hybrid C-O-Ne white dwarfs be progenitors of SNe Ia?}

To answer this question, we allow our naked hybrid white dwarfs from the left panel of Fig.~2 to accrete material 
with chemical compositions equal to the ones of their C-rich surface buffer zones. Our used accretion rate, 
$\dot{M}_\mathrm{acc} = 8\times 10^{-7} M_\odot/\mathrm{yr}$, is sufficiently high for H accreted with 
the same rate to be burnt into He 
under stationary conditions. We assume that He is then transformed into C in a He shell burning. 
We take into account compressional heating of the accreted material, but neglect energy generation 
in the H and He burning. Our main goal is to find out what will happen with our hybrid white dwarfs 
when their masses approach the Chandrasekhar limit.


\section{The Urca process uncertainty}

This is one of the most important problems for our hybrid SNIa progenitor models, 
because C burning results in relatively high abundances of $^{25}$Mg and $^{23}$Na (Fig.~2, right panel) 
that form Urca pairs with $^{25}$Na and $^{23}$Ne, respectively. The Urca reactions involving 
the first and second pair are activated at the densities $1.29\times 10^9 \mathrm{g}/\mathrm{cm}^3$ 
and $1.66\times 10^9 \mathrm{g}/\mathrm{cm}^3$, 
both of which are below the central density of explosive C ignition. Therefore, 
we anticipate a strong effect of these Urca processes on the final thermal and chemical 
structures of our SNIa progenitor models, especially in the presence of convection, 
because the kinetic energy of convective motion serves as a source for both Urca cooling 
and heating, which is very difficult to take into account \cite{bk:01}. In these circumstances, 
we choose to present the results of our calculations of SNIa progenitor models 
for a number of different mixing assumptions that mimic potentially possible outcomes of 
the interaction of convection and Urca reactions. Convective Urca heating and cooling are 
only included in our calculations of the nuclear energy generation rate, for which we use  
the most recent well-resolved Urca reaction and neutrino energy loss data of 
Toki et al. \cite{toki:13}, but they are ignored in our application of the mixing length theory.

\begin{figure}
\includegraphics[width=.5\textwidth]{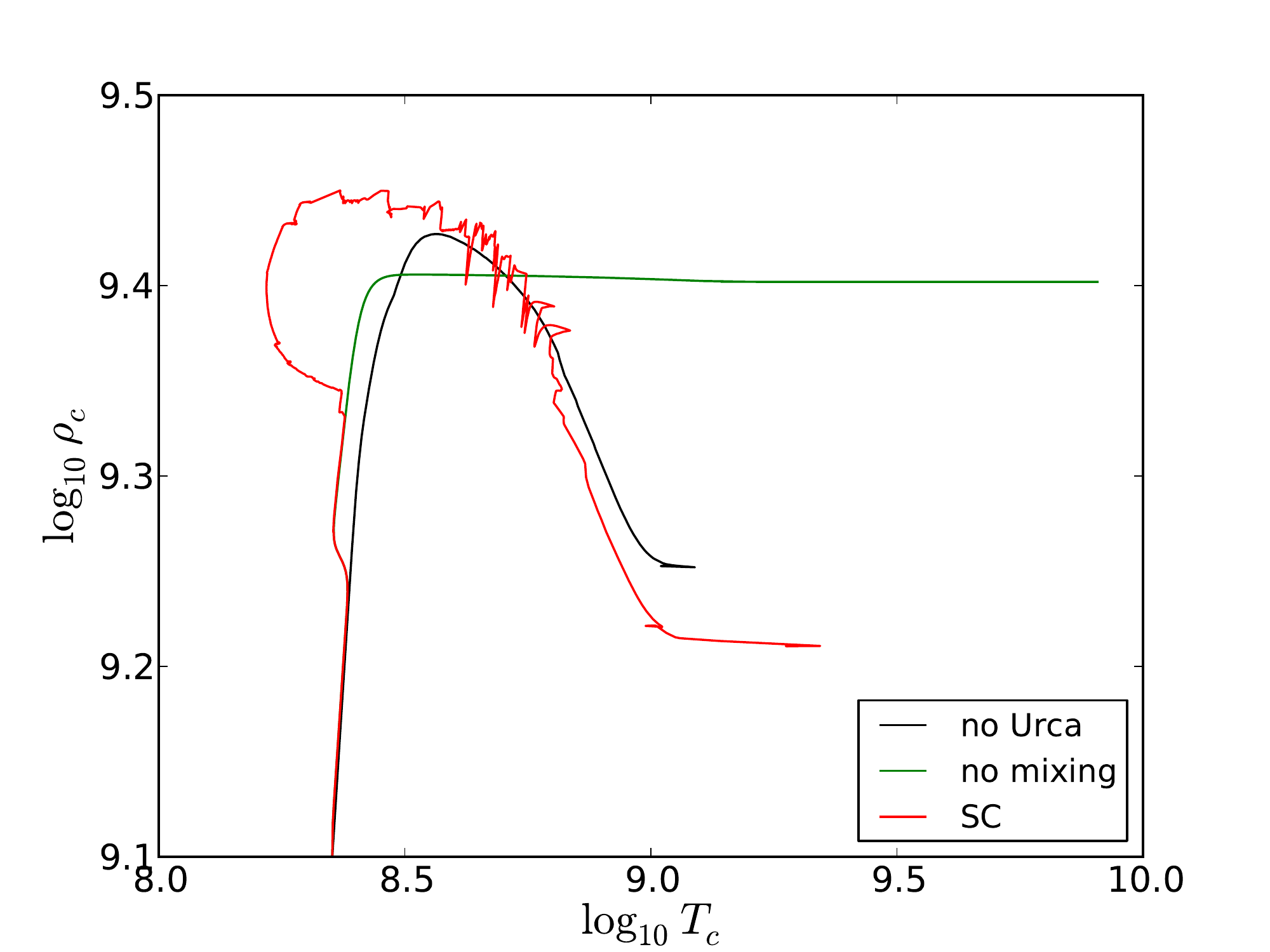}
\includegraphics[width=.5\textwidth]{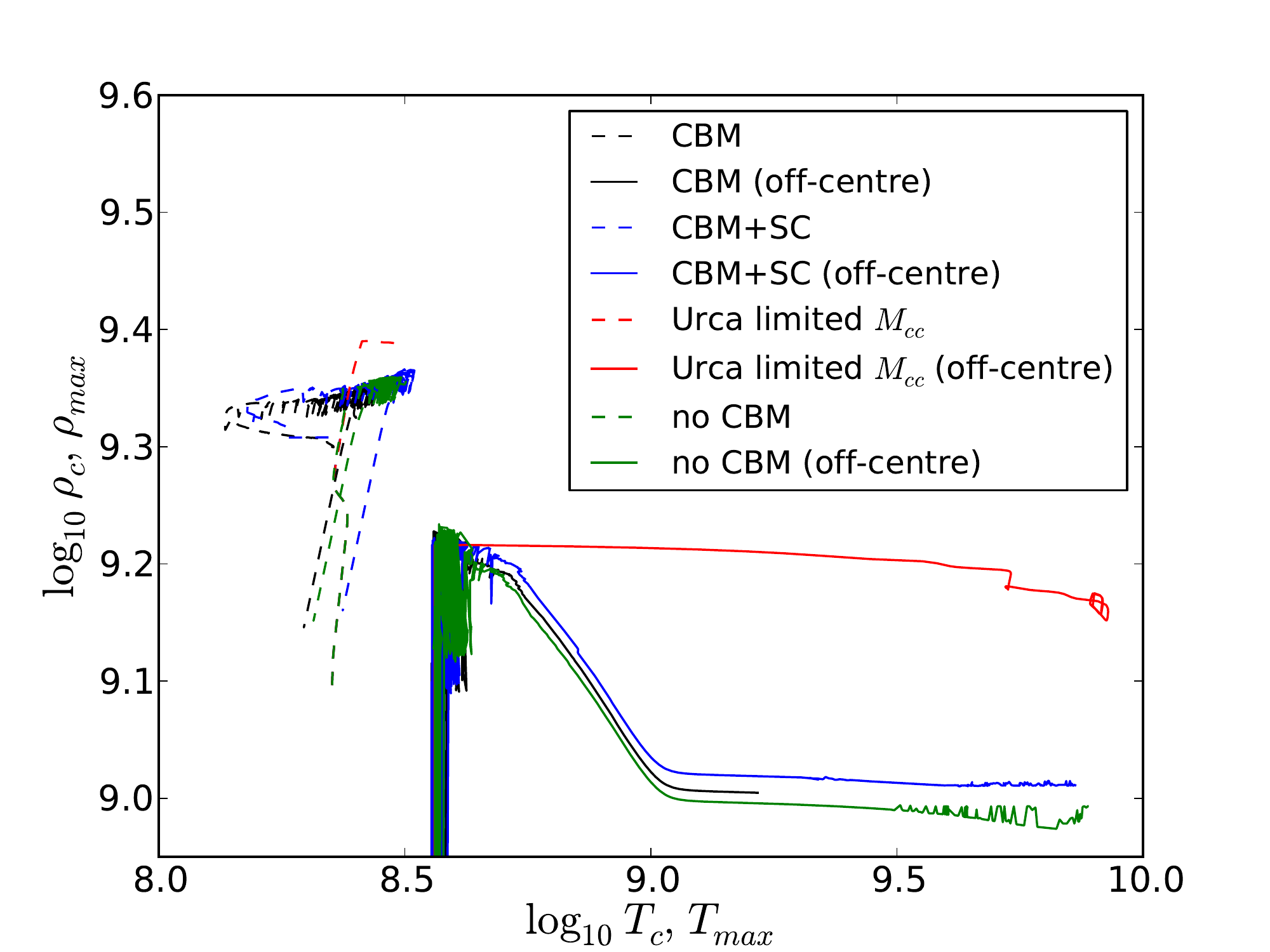}
\caption{{\bf Left panel:} the evolution of central temperature and density resulting from  
the accretion of C-rich surface-composition material onto the hybrid C-O-Ne white dwarf with 
the rate of $8\times 10^{-7} M_\odot/\mathrm{yr}$. The initial mass of the star is $6.9 M_\odot$. The simulations have been done 
for the following mixing assumptions: no Urca processes are included (no Urca); convective mixing is 
totally suppressed by Urca processes (no mixing); besides convection, only semiconvection is included, 
but convective boundary mixing is neglected (SC). In these three cases, the explosive C ignition occurs in the center.
{\bf Right panel:} same as in the left panel but, because in these cases the explosive C ignition occurs off center, 
the evolution of the maximum temperature (off center) is also plotted. The simulations have been done 
for the following assumptions: only convective boundary mixing is included as an extra mixing process (CBM, Fig.~4); 
both CBM and SC are included (CBM+SC); Urca processes suppress convection between the center and 
the $^{23}$Na/$^{23}$Ne Urca shell (Urca limited $M_\mathrm{cc}$); only convective mixing is included (no CBM).
}
\label{fig3}
\end{figure}

\section{Results}

For the initial masses between $6.5 M_\odot$ and $6.9 M_\odot$, we obtain hybrid C-O-Ne white dwarfs 
that can be progenitors of SNe Ia. However, because of the Urca process uncertainty, 
their thermal and chemical structures at the moment of explosive C ignition remain uncertain. 
Depending on our assumption about the interaction between convection and Urca reactions, 
the explosive C ignition occurs either in or off the center (left and right panels of Fig.~3). 
Because of this uncertainty, which can only be resolved in future reactive-convective 3D 
hydrodynamical simulations, we have generated several potentially possible instances of 
the hybrid SN Ia progenitor model that can be used, in the meanwhile, as initial models 
for 2D or 3D simulations of SN Ia explosion. Their resulting nucleosynthesis yields and 
light curves can be compared with available SN Ia observations to find out which of 
the instances better fits the observational data. 

\begin{figure}
\begin{center}
\includegraphics[width=.8\textwidth]{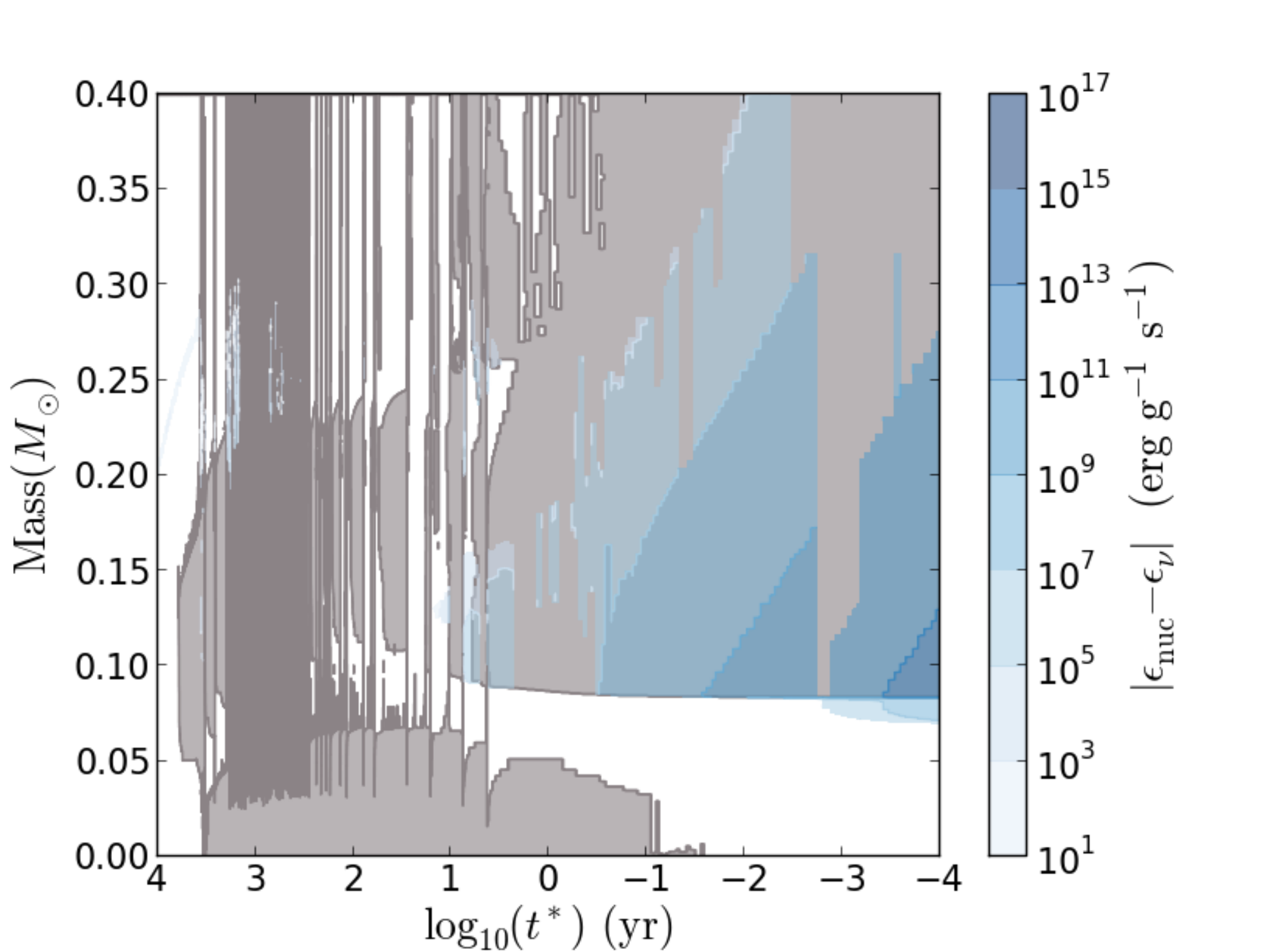}
\end{center}
\caption{Zoomed in Kippenhahn diagram for the CBM model from the right panel of Fig.~3 
when it approaches the explosive C ignition. The blue shades show the energy generation 
by the Urca reactions and C burning.
}
\label{fig4}
\end{figure}

\section{Conclusion}

Although it can be expected that the less massive of our stars will enter a phase of 
explosive C-burning once their cores approach the Chandrasekhar mass, it is unlikely 
that the outcome will be a normal SN Ia. If the burning would proceed as a detonation 
from the start, the lack of intermediate-mass elements in the ejecta would, 
like in the pure CO white dwarfs, contradict observations. In contrast, a deflagration 
ignited in the CO core cannot change into a detonation in the ONe zone easily because 
the critical mass for a detonation of an O and Ne mixture is much larger than that of carbon. 
Therefore, the more likely outcome of such an explosion is a faint SN Ia, similar to the SN 2002cx class, 
as was also found for pure-deflagration CO white dwarfs, and possibly a bound remnant is 
left behind \cite{fink:14}.
Estimates of SNIa birthrates with our hybrid WD
progenitors have recently been made in \cite{meng:14} and \cite{wang:14}.

\medskip

{\bf Acknowledgments.} This research has been supported by the National Science Foundation under grants 
PHY 11-25915 and AST 11-09174. This project was also supported by JINA (NSF grant PHY 08-22648). 
Falk Herwig acknowledges funding from Natural Sciences and Engineering Research Council of 
Canada (NSERC) through a Discovery Grant. Pavel Denissenkov thanks Wolfgang Hillebrandt for his comments. 
Ken Nomoto acknowledges the support from the Grant-in-Aid for Scientific Research 
(23224004, 23540262, 26400222) from the Japan Society for the Promotion of Science, 
and the World Premier International Research Center Initiative (WPI Initiative), MEXT, Japan.

\end{document}